\documentclass[preprint,revtex4]{emulateapj}
\usepackage{apjfonts}
\usepackage{xspace}
\usepackage[colorlinks=true,linkcolor=blue,citecolor=blue]{hyperref}

\newcommand{\ml}{\ensuremath{\Msun\,\mathrm{yr}^{-1}}}
\newcommand{\Msun}{\ensuremath{M_\odot}}
\newcommand{\kms}{\ensuremath{\mathrm{km\,s}^{-1}}}

\begin{document}

\title{Searching for the expelled hydrogen envelope in Type I supernovae \\
via late-time H$\alpha$ emission}

\author{J. Vinko\altaffilmark{1,2,3}, D. Pooley\altaffilmark{4}, 
J. M. Silverman\altaffilmark{1}, J. C. Wheeler\altaffilmark{1}, T. Szalai\altaffilmark{3}, 
P. Kelly\altaffilmark{5}, P. MacQueen\altaffilmark{1}, G. H. Marion\altaffilmark{1} \\
and K. S\'arneczky\altaffilmark{2}}

\altaffiltext{1}{Department of Astronomy, University of Texas at Austin, Austin, TX, 78712, USA}
\altaffiltext{2}{Konkoly Observatory, Research Centre for Astronomy and Earth Sciences, Hungarian Academy of Sciences, Konkoly Thege ut 15-17, Budapest, 1121, Hungary}
\altaffiltext{3}{Department of Optics \& Quantum Electronics, University of Szeged, Dom ter 9, Szeged 6720, Hungary}
\altaffiltext{4}{Department of Physics and Astronomy, Trinity University, One Trinity Place, San Antonio, TX, 78212, USA}
\altaffiltext{5}{Department of Astronomy, University of California at Berkeley, 501 Campbell Hall, Berkeley, CA, 94720-3411, USA}

\begin{abstract}
We report the first results from our long-term observational survey aimed at
discovering late-time interaction between the ejecta 
of hydrogen-poor Type I supernovae and the hydrogen-rich envelope expelled from 
the progenitor star several decades/centuries before explosion. The expelled envelope,
moving with a velocity of $\sim 10$--$100$ km s$^{-1}$, is expected to be caught up
by the fast-moving SN ejecta several years/decades after explosion depending on the
history of the mass-loss process acting in the progenitor star prior to explosion. 
The collision between the SN ejecta and the circumstellar envelope results in 
net emission in the Balmer-lines, especially in H$\alpha$. We look for signs of late-time H$\alpha$
emission in older Type Ia/Ibc/IIb SNe having hydrogen-poor ejecta, via
narrow-band imaging. Continuum-subtracted H$\alpha$ emission has been detected for 13 point 
sources: 9 SN Ibc, 1 SN IIb and 3 SN Ia events. Thirty-eight SN sites were observed on at least 
two epochs, from which
three objects (SN~1985F, SN~2005kl, SN~2012fh) showed significant temporal variation 
in the strength of their H$\alpha$ emission in our DIAFI data. 
This suggests that the variable emission is probably not due to nearby H~II regions 
unassociated with the SN, 
and hence is an important additional hint that ejecta-CSM interaction 
may take place in these systems. Moreover, we successfully detected the
late-time H$\alpha$ emission from the Type Ib SN~2014C, which was recently
discovered as a strongly interacting SN in various (radio, infrared, optical and X-ray) bands.
\end{abstract}

\keywords{(stars:) supernovae: general; stars: winds, outflows; (ISM:) H~II regions }

\section{Introduction} \label{sec:intro}

All stars are thought to lose mass during their main-sequence and subsequent 
evolution.  This can range from a paltry stellar wind like that from the Sun 
($10^{-14}$~\ml\ with speeds of hundreds of \kms) to the slow, dense winds of 
red giants ($10^{-8}$--$10^{-6}$~\ml with speeds of tens of \kms) to brief, 
violent, $\eta$~Carinae-like expulsions of tens of solar masses from very massive 
stars.   At different stages of evolution, a star will undoubtedly have different 
modes of mass loss. Binary evolution will play a role in many cases. 

While it is impossible to study in real time these different 
modes of mass loss in an individual star, we can do so by studying 
individual supernovae (SNe) for years and even decades after the explosion.  The 
fast-moving SN shock is effectively a time machine, encountering material shed 
earlier in the life of the pre-supernova star.

Recent years have seen a new focus on the circumstellar media (CSM) 
surrounding SN sites and the interaction of supernova ejecta and shocks with CSM.  
Well-known CSM-interacting SN types are the Type IIP \citep{chugai07}, which arise 
in red supergiant progenitors that blow nearly steady-state winds into which the 
star explodes and the Type IIn (SN~IIn) events, which have long been recognized 
to show narrow emission lines that reveal dense CSM \citep[e.g.][]{alex97}. 
In addition, the great luminosities
of superluminous supernovae (SLSNe) are also thought to be powered, in some cases, 
by ejecta-CSM interaction \citep{manos13}. 

The various types of stripped-envelope SNe are also of great interest in this respect,
because these events have little or no hydrogen on the progenitor at the time of explosion, 
but the progenitors must have once been normal hydrogen-rich stars. Such events include the
Type IIb SNe, in which the hydrogen envelope has been partly expelled prior to explosion,
as well as the Type Ibc, which essentially lost their H-rich envelope. The mechanism
of these processes is ill-understood: it could be due to winds \citep{heger97, puls08}, 
episodic ejection \citep{smith06, pastor07, shiqua14} or binary 
interactions, including common-envelope formation and ejection \citep[see the reviews by][]{taam00, smith14}. 
Even some Type Ia SNe, referred to as SNe Ia-CSM, show strong 
H$\alpha$ emission, a clear sign of CSM-interaction, in their late-phase optical spectra 
\citep{silverman13a, inserra16}.  

The collision between the fast-moving SN ejecta and the slow-moving CSM creates the 
well-known double-shock pattern having the forward shock (FS) propagating into the CSM and
the reverse shock {\bf (RS)} propagating back into the ejecta \citep[e.g.][and references therein]{cf03}.
The acceleration of free electrons results in strong non-thermal emission, from X-rays to
radio, coming from the interaction site. Massive SN progenitors ($M \gtrsim 20$ M$_\odot$) 
can produce strong, fast-moving winds prior to explosion, which can create a
low-density cavity around the explosion site, surrounded by a relatively dense CSM shell 
consisting of previously expelled (H-rich) material \citep[e.g.][]{cl89}. 
When the SN ejecta hits this dense shell, it drives a strong RS into the ejecta, 
which leads to strong X-ray radiation from the region between the RS and the ejecta-CSM interface 
\citep{nymark06}. The hard X-rays produced by the RS are mostly absorbed by the cool, dense CSM 
shell causing strong ionization in this medium. Subsequent recombination in this H-rich CSM shell 
results in emergent emission in the hydrogen Balmer-lines, mostly H$\alpha$ \citep{cc06}.
This mechanism is thought to produce intermediate-width ($FWMH \sim 2000$ - $3000$ km~s$^{-1}$) 
emission lines, mostly due to multiple scattering on free electrons within the shell and, to a lesser 
extent, bulk motion caused by the acceleration of the shell by the expanding SN ejecta at the beginning
of the interaction phase. 
The FS propagating into the CSM can also ionize the surrounding material, which may produce additional, 
narrow ($FWHM \lesssim 100$ km~s$^{-1}$) H$\alpha$ emission coming from the pre-shock CSM in front of the FS.
The appearance of the Balmer-emission can be especially interesting in the case of stripped-envelope
SNe, since in such cases the ejecta contain only very low or negligible amount of H. Thus,
the emerging Balmer-emission is a very strong indication that the ejecta have overrun the cavity
and plunged into the H-rich CSM shell.

The natural expectation that the fast-moving SN ejecta must overtake the previously expelled
H-rich envelope has been beautifully demonstrated in the case of the Type Ibc SN~2001em,
where strong radio \citep{stock04} and X-ray emission \citep{pooley04} was discovered $\sim 3$
years after explosion. These discoveries generated further interest in SN~2001em, because
the radio luminosity ($L_\mathrm{radio} \approx 2 \times 10^{28}$ erg s$^{-1}$ Hz$^{-1}$ at 6 cm) 
as well as the X-ray luminosity ($L_\mathrm{X} \approx 10^{41}$ erg s$^{-1}$) were far above anything 
seen from other SNe Ibc at an age of several years. Since some fraction of SN Ibc give rise to gamma-ray bursts (GRB) where the jet is aimed in our direction, it was hypothesized by \citet{granot04} 
that SN~2001em may have been an off-axis GRB. \citet{pooley04}, however, suggested the strong interaction
between the SN ejecta and a dense CSM as an alternative mechanism which also can produce the
exceptionally strong X-ray luminosity as observed by {\it Chandra}.  Indeed, the late-time optical 
spectrum of SN~2001em, which revealed strong H$\alpha$ emission \citep{soderberg04}, added
strong support to the ejecta-CSM interaction scenario. 
\citet{cc06} explained all the unusual late-time X-ray, radio, and optical properties, as well as the failure to resolve a possible jet via VLBI observations \citep{bieten05, schinzel09}, by suggesting that the SN ejecta had finally caught up with the cast-off hydrogen envelope of the progenitor star and that 
strong interaction was taking place. 

A similar phenomenon is invoked in the case of PTF~11kx \citep{dilday12, silverman13b}, a Type Ia-CSM, where strong H$\alpha$ emission developed $\sim 40$d after maximum in an otherwise normal-looking Type Ia SN.

Motivated by the examples above, we started an observational survey to monitor several 
years-to-decades-old H-deficient SNe in order to catch signs of the starting (or ongoing)
ejecta-CSM interaction. Our concept is that numerous SN Ia and/or Ibc may show 
detectable CSM interaction at such late phases, because their CSM shells may have 
reached greater distances from the explosion site than in those well-known cases when the
interaction started within a few hundred days after explosion.
Assuming that the H-rich envelope travels with a speed between 10 and 100 km~s$^{-1}$ 
\citep{cc06}, the interaction with the fast-moving ejecta ($v_{SN} \sim 10,000$ km~s$^{-1}$)
is expected to start roughly a decade after SN explosion if the expulsion of the envelope ended
$\sim 10^4$ -- $10^3$ years before core collapse, depending on wind speed.

The aim of our observations is the direct detection of the
H$\alpha$ emission due to the collision between the H-poor SN ejecta and the surrounding
H-rich CSM via imaging the SN sites through narrow-band filters centered on the (redshifted)
wavelength of H$\alpha$. In Sections~\ref{sec:obs} and \ref{sec:method} we give details 
of the observations and our methodology of detecting H$\alpha$ emissions. 
Section~\ref{sec:res} shows our early results, which are further discussed in 
Section~\ref{sec:discuss}. Section~\ref{sec:concl} summarizes our conclusions.

\section{Observations} \label{sec:obs}

\begin{figure}[ht!]
\figurenum{1}
\plotone{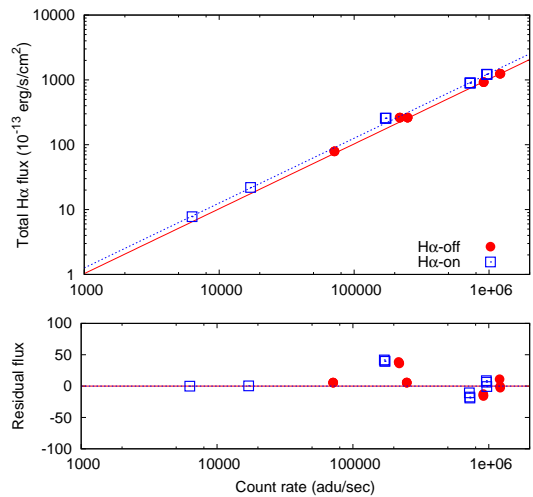}
\caption{Flux calibrations for the H$\alpha$-on (blue squares) and H$\alpha$-off (red circles) 
filters.
\label{fig:f1}}
\end{figure}

We utilized the 2.7m (107") Harlan J. Smith Telescope equipped with the 
Direct Imaging Auxiliary Functions Instrument (DIAFI) at McDonald Observatory
for conducting the imaging survey. We applied a
narrow-band ($FWHM=7$ nm) H$\alpha$ filter centered on the redshift of the 
Virgo-cluster ($z =0.0033$, $\lambda = 6585$ \AA) and another similar filter at an 
off-line position ($\lambda = 6675$ \AA) as a guard-band filter to measure
the continuum flux near the H$\alpha$ line. In the following we refer to
these filters as H$\alpha$-on and H$\alpha$-off, respectively.  
 
We have selected a subsample of the full list of 3662 known SN~I (SN~Ia, SN~Ibc) and SN~IIb 
discovered before 2014, based on the following selection criteria: 
\begin{itemize}
\item{declination higher than $-30 \deg$;}
\item{distance less than $200$ Mpc.}
\end{itemize} 
This resulted in 747 potential candidates. These are sampled further,
concentrating on the closest ($D \lesssim 30$ Mpc) events. The number of SNe
in the restricted sample is 178. 

Beside observability from the northern hemisphere, the reason for these selection
criteria were twofold: $i)$ to have the H$\alpha$ line redshifted into the 
transmission band of the H$\alpha$-on filter, and $ii)$ to be able to reach 
a signal-to-noise ratio of $\sim 50$ for $L(\mathrm{H}\alpha) \sim 10^{39}$ erg~s$^{-1}$
(i.e. a signal similar to SN~2001em) within $\sim 1$ hour exposure time with DIAFI.

We started our multi-season observing campaign in 2014, using both the
H$\alpha$-on and -off filters on the SNe with $D < 30$ Mpc. Since
then we have imaged 76 galaxies hosting 99 SNe, i.e. $\sim 55$ percent of the total
sample. 
The journal of the observations can be found in Table~\ref{tab:journal}.

\begin{table}
\begin{center}
\caption{The journal of narrow-band imaging observations with DIAFI}
\begin{tabular}{lcc}\label{tab:journal}
Start date & End date & No. of observed SN hosts\\
\hline
2014 Feb 27 & 2014 Feb 28 & 22 \\
2014 May 03 & 2014 May 04 & 17 \\
2014 Sep 30 & 2014 Oct 02 & 23 \\
2015 Mar 14 & 2015 Mar 16 & 27 \\
2015 May 20 & 2015 May 20 & 7 \\
2015 Aug 23 & 2015 Aug 23 & 4 \\
2016 Jun 07 & 2016 Jun 10 & 17 \\
\hline
\end{tabular}
\end{center}
\end{table}

The H$\alpha$ emission produced by SN-CSM interaction is likely to be variable in time.
This offers an opportunity to distinguish between the "real" interaction-produced
emission and the flux coming from the immediate vicinity of the SN site but not related
to the SN blast wave (e.g. from a nearby H~II cloud), because the latter sources are expected
to remain roughly constant in time. Thus, instead of maximizing the number of observed SNe,
we aimed at taking multiple images of those SNe that were detected as H$\alpha$ emitters
at significantly different epochs. This effort resulted in 38 SNe (in 32 different host
galaxies) imaged at least twice, separated by a 3-month-long or longer temporal baseline.

\section{Method} \label{sec:method}

\begin{figure*}
\figurenum{2}
\begin{center}
\includegraphics[height=10cm]{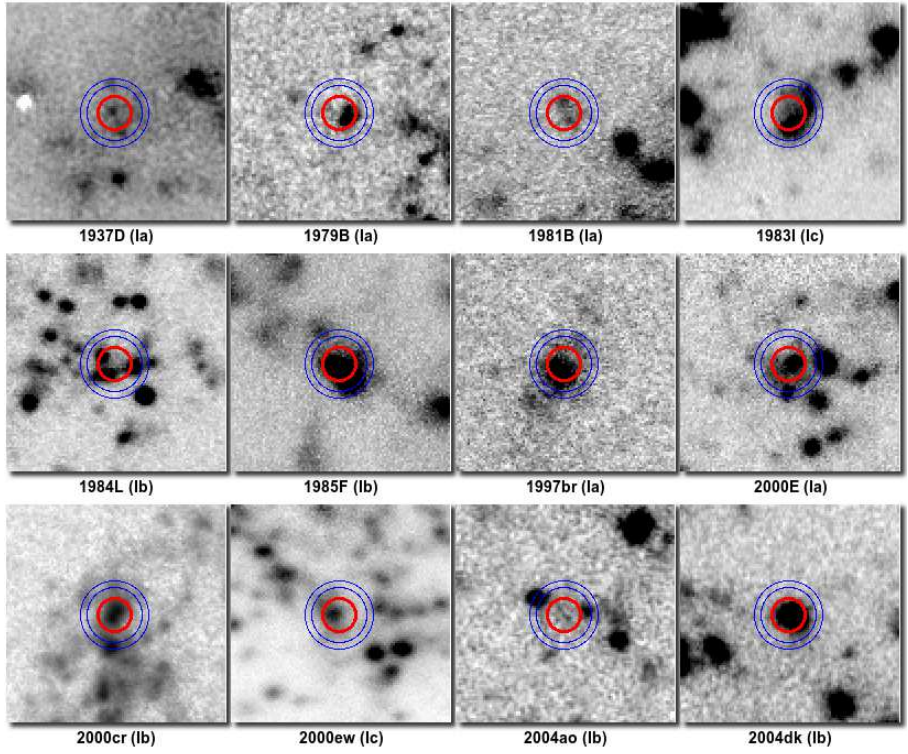}
\end{center}
\caption{Detections of significant (S/N $>$ 4) H$\alpha$ emission from the sample SN sites.
Each stamp is a $0.7 \times 0.7$ arcmin$^2$ subset of the continuum-subtracted DIAFI
frame centered on the SN position. The photometric aperture and annulus are
indicated by the red and blue circles, respectively.\label{fig:f2}}
\end{figure*}

\begin{figure*}
\figurenum{3}
\begin{center}
\includegraphics[height=10cm]{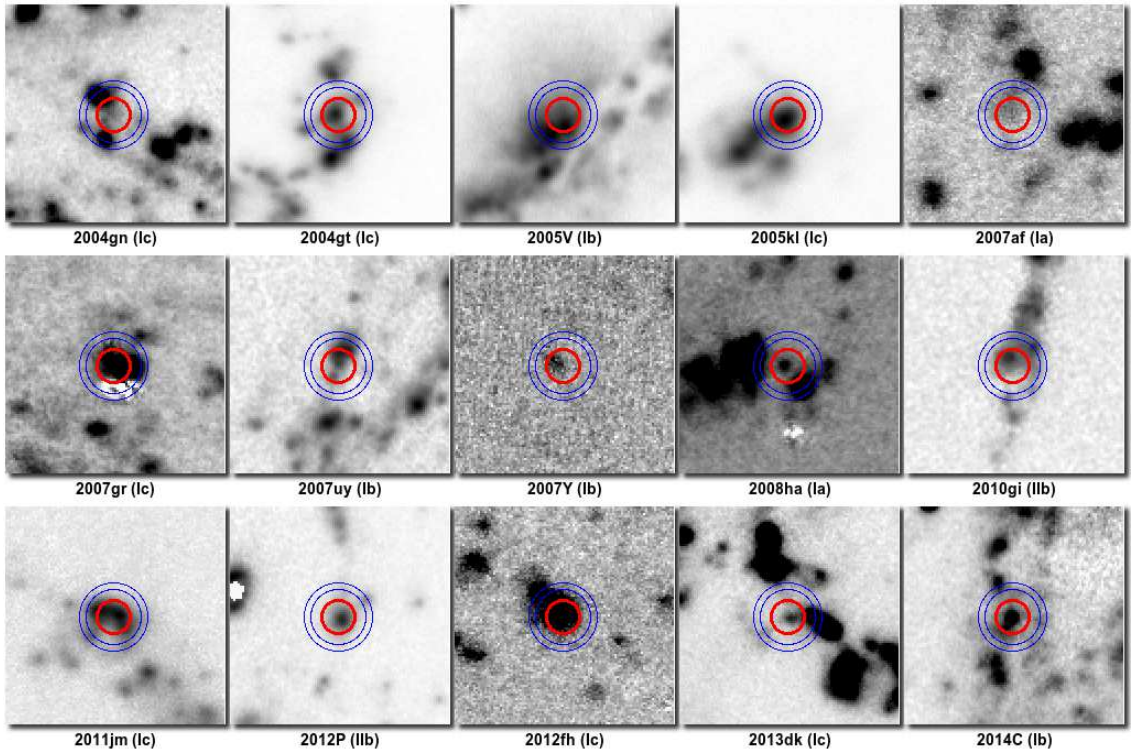}
\end{center}
\caption{Same as Fig.\ref{fig:f2}, but for additional SNe.\label{fig:f3}}
\end{figure*}

The frames taken with DIAFI were reduced and calibrated in the following way. 
Three frames per filter were collected for each SNe.   
Following standard bias-subtraction and flatfield-division, the frames obtained through the
same filter were geometrically registered by applying the 
{\it IRAF}\footnote{Image Reduction and Analysis Facility (IRAF) 
is distributed by the National Optical Astronomy Observatories,
which are operated by the Association of Universities for Research
in Astronomy, Inc., under cooperative agreement with the National Science Foundation.} 
task {\it xregister}, then 
they were median-combined to filter out the numerous cosmic-ray hits. 
The median-combined frames were then transformed to the World Coordinate System (WCS) 
by using the
{\it SEXtractor}\footnote{http://www.astromatic.net/software/sextractor} and 
{\it WCSTools}\footnote{http://http://tdc-www.harvard.edu/wcstools/} codes. 
Having the combined images properly registered in both filters, the
{\it HOTPANTS}\footnote{http://www.astro.washington.edu/users/becker/v2.0/hotpants.html}
code was applied to subtract the H$\alpha$-off frames from the H$\alpha$-on frames to
produce the continuum-subtracted difference images. These images were used to look for
any H$\alpha$-emitting source in the immediate vicinity of the SN as described below.

Flux calibration was performed via observing spectroscopic flux standard stars. We measured the
observed count rates on the median-combined H$\alpha$-on and -off frames via aperture photometry 
using the {\it YODA}\footnote{http://www.as.utexas.edu/$\sim$drory/yoda/index.html} ("Yet anOther object Detection
Application") code \citep{drory03}. 
The diameter of the circular aperture was chosen as $d_{ap} =15$ pixels ($\sim 6.1$ arcsec),
and the fluxes within this aperture were integrated. The subtracted background flux was estimated 
within an annulus having inner diameter and width of $20$ and $5$ pixels, respectively, 
around each object. The correction for the atmospheric extinction was computed using the 
wavelength-dependent KPNO extincion function as tabulated in {\it IRAF}, although this was found
negligible compared to other uncertainties. 
For the observed spectroscopic standard stars, the true H$\alpha$ fluxes were
calculated as the integral of their known spectral flux densities within the spectral bandpasses
of the applied filters. Finally, the conversion between the observed count rates ($CR$, in ADU/s) and the
true H$\alpha$ fluxes for both filters have been determined by fitting the following simple relation to
the data: 
\begin{equation}
F_{H\alpha} \times 10^{13} \mathrm{(erg/s/cm^2)} ~=~ K \cdot CR \mathrm{(ADU/s)}
\label{eq-1}
\end{equation}

Figure~\ref{fig:f1} illustrates the quality of the fit. The $K$ scaling factors obtained
this way turned out to be $1.03 \times 10^{-3}$ and $1.26 \times 10^{-3}$ for the 
H$\alpha$-off and -on filters, respectively. The uncertainties of the flux calibration,
estimated from the scatter of the residual plots in the bottom panel of Figure~\ref{fig:f1},
are $1.89 \times 10^{-2}$ and $2.38 ~\times 10^{-2}$ erg s$^{-1}$ cm$^{-2}$, respectively.
  
The continuum-subtracted H$\alpha$ flux at the position of every target SNe was estimated
by the combination of two approaches.  First, aperture photometry (using the same parameters 
as above) was computed with {\it YODA} at the SN position on the continuum-subtracted difference frames
produced by {\it HOTPANTS}.  
If the total flux in the aperture exceeded the local sky level, the signal-to-noise ratio (S/N) was
estimated by dividing the total flux by its uncertainty reported by {\it YODA}. Objects having
S/N $>$ 4 were selected as detection candidates. Second, the detection candidates were examined
by computing the same aperture photometry as before, but on the final, median-combined H$\alpha$-on
and H$\alpha$-off frames. After transforming the measured count rates to physical fluxes (applying
Equation~\ref{eq-1}), the net H$\alpha$ flux was computed by subtracting the calibrated fluxes from the
H$\alpha$-off frames from the fluxes measured on the H$\alpha$-on images. Again, uncertainties are 
estimated by adopting the values as reported by {\it YODA}. Only those candidates that showed 
S/N $>$ 4 after this step were retained as detections. This way it was possible to filter out some bogus
detections, which were due to numerical artifacts on the difference images given by {\it HOTPANTS}.

\begin{figure}[ht!]
\figurenum{4}
\includegraphics[width=8.5cm]{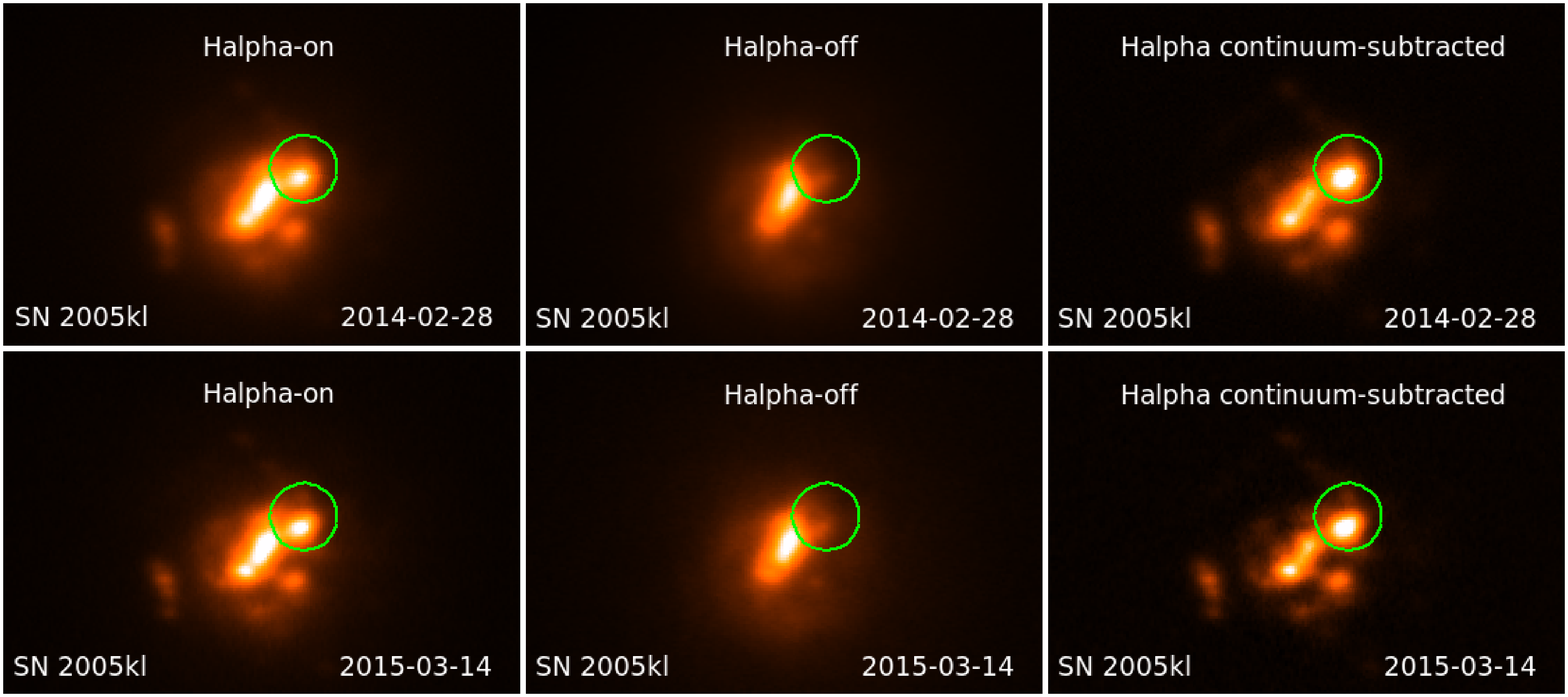}
\caption{Illustration of the methodology in the case of SN 2005kl, which showed
the strongest H$\alpha$ flux among the detected objects. The first column shows the two frames 
taken through the H$\alpha$-on filter on two different epochs, the middle column contains the
two H$\alpha$-off frames, while the third column displays the continuum-subtracted (i.e. H$\alpha$-on
minus H$\alpha$-off) frames. The size of the frames is $1.6 \times 1.1$ arcmin$^2$. 
The position of the SN is encircled. \label{fig:f4}}
\end{figure}

For those objects that were observed on more than one epoch, the effect of different atmospheric 
transparency was corrected for in the following way: photometry of bright, unsaturated stellar sources
on the H$\alpha$-off (continuum) frames were computed, and their instrumental magnitudes 
obtained on the 2nd night were scaled to those taken on the 1st night. Similar transformation 
was then computed for the H$\alpha$-on frames as well. This resulted in consistent relative photometry
on the two datasets within an error of a few percent. These data were used to construct the 
H$\alpha$ "light curves" to look for temporal variability among the detected sources 
(see Section~\ref{sec:res}).

\section{Results}\label{sec:res}

\begin{table}
\begin{center}
\caption{List of detected objects and their H$\alpha$ luminosities 
(computed in erg s$^{-1}$
using the distances in the 3rd column). See text for details.}
\begin{tabular}{lccccl}\label{table:detect}
SN & Type & D & $\log \mathrm{L}(\mathrm{H}\alpha)$ & unc. & Remark \\
\hline
 & & (Mpc) & (cgs) & (dex) & \\
\hline
1937D & Ia & 11.0 & 37.8135 & 0.0277 & point source \\
1979B & Ia & 17.0 & 37.9868 & 0.0070 & diffuse \\
1981B & Ia & 17.7 & 37.7577 & 0.0868 & diffuse \\
1983I & Ic & 14.0 & 38.6929 & 0.0048 & diffuse \\
1984L & Ib & 18.8 & 38.5064 & 0.0341 & diffuse \\
1985F & Ib & 7.6 & 38.6909 & 0.0121 & point source \\
1997br & Ia & 27.5 & 38.3376 & 0.0374 & diffuse \\
2000E & Ia & 23.9 & 38.8246 & 0.0231 & point source \\
2000cr & Ib & 57.6 & 39.5134 & 0.0272 & point source \\
2000ew & Ic & 17.6 & 39.6804 & 0.0647 & point source \\
2004ao & Ib & 26.9 & 38.0196 & 0.0664 & diffuse \\
2004dk & Ib & 22.8 & 39.1888 & 0.0079 & point source \\
2004gn & Ic & 14.4 & 38.2218 & 0.1629 & diffuse \\
2004gt & Ic & 21.4 & 40.0318 & 0.0109 & point source \\
2005V & Ib & 22.4 & 40.3510 & 0.0095 & diffuse \\ 
2005kl & Ic & 21.6 & 41.8452 & 0.0020 & point source \\
2007Y & Ib & 19.1 & 37.8156 & 0.0272 & diffuse \\
2007af & Ia & 24.3 & 37.8710 & 0.0676 & diffuse \\
2007gr & Ic & 8.2 &  38.4519 & 0.0120 & diffuse \\
2007uy & Ib & 30.0 & 39.2632 & 0.0115 & diffuse \\
2008ha & Ia & 20.0 & 38.2445 & 0.0684 & point source \\
2010gi & IIb & 10.2 & 38.4090 & 0.0635 & diffuse \\
2011jm & Ic & 22.2 & 39.5923 & 0.0030 & diffuse \\
2012fh & Ic & 16.4 & 38.8490 & 0.0008 & point source \\
2012P & IIb & 25.7 & 39.4327 & 0.0070 & point source \\
2013dk & Ic & 21.4 & 38.8910 & 0.0799 & point source \\
2014C & Ib & 14.3 & 38.8262 & 0.0527 & point source \\
\hline
\end{tabular}
\end{center}
\end{table}

We have detected (using the criterion of S/N $> 4$) 
continuum-subtracted H$\alpha$ emission from 27 SN sites (see Table~\ref{table:detect}).
At least 14 of these detections
($\sim 50$ percent) are clearly from diffuse H$\alpha$ emitting sources, probably
nearby H~II areas. The others look like point sources in our DIAFI frames, but 
some of them appear slightly off-center with respect to the expected SN position,
and thus could just be nearby compact H~II clouds. Figure~\ref{fig:f2} and \ref{fig:f3}
show the collection of the positive detections.

\begin{figure}[ht!]
\figurenum{5}
\plotone{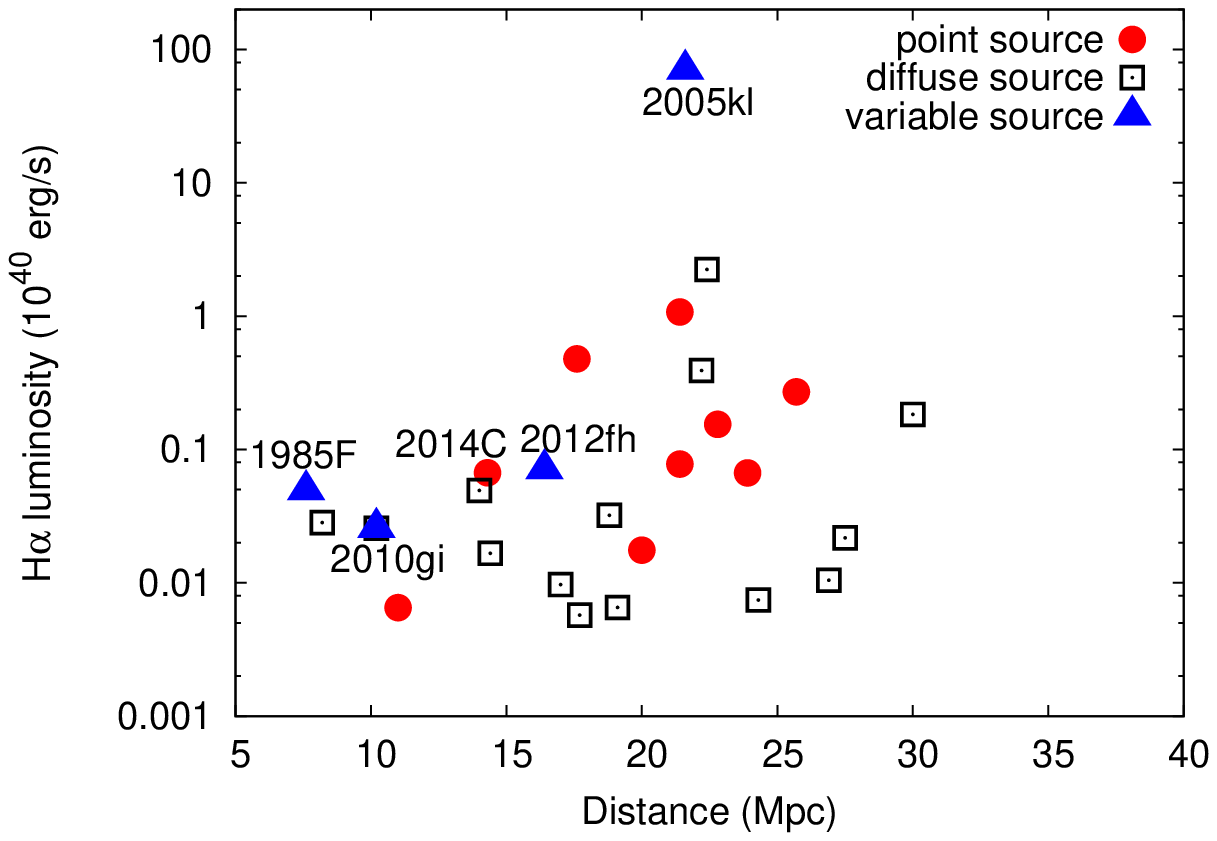}
\caption{Derived H$\alpha$ luminosities of the detected sources plotted against distance. 
Filled symbols represent the point-like sources on the DIAFI images,
while open squares denote more extended (diffuse) emitting sources (see Table~\ref{table:detect}).
Sources that show significant temporal variability (Figure~~\ref{fig:f6}) are plotted with
triangles.
\label{fig:f5}}
\end{figure}

The 3rd column of Table~\ref{table:detect} contains the average of the redshift-independent distances
taken from the NASA Extragalactic Database\footnote{http://ned.ipac.caltech.edu} (NED).
Based on these distances and the Milky Way extinction coefficient in the $R$-band from \citet{sf11},
the logarithm of the calculated H$\alpha$ luminosities and their uncertainties (expressed in
erg s$^{-1}$) are shown in the 4th and 5th
columns of Table~\ref{table:detect}. The (likely underestimated) uncertainties are pure photometric
errors described above and do not contain any other effects, e.g. the uncertainty of
the distances or any in-host extinction.

\begin{figure}[ht!]
\figurenum{6}
\plotone{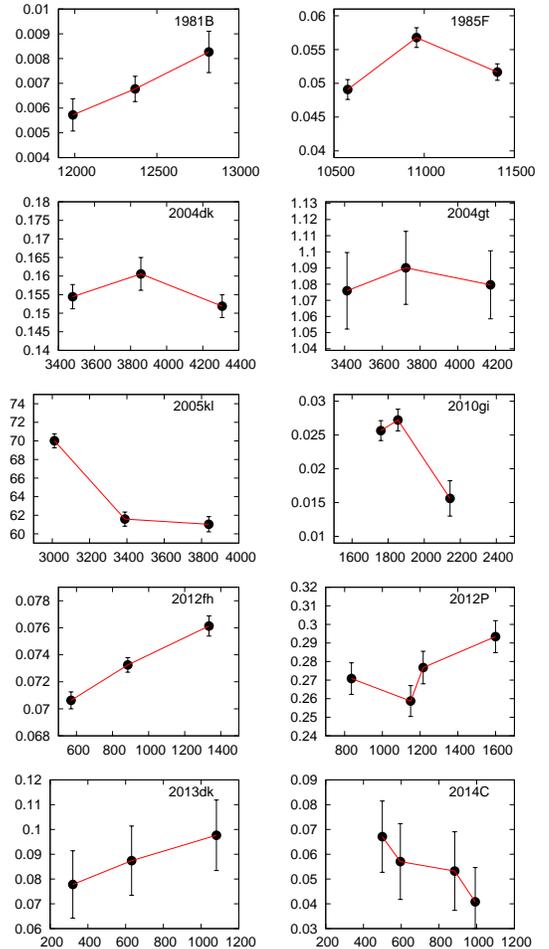}
\caption{Variation of the detected H$\alpha$ luminosities. Each panel shows the  
H$\alpha$ luminosity (in $10^{40}$ erg s$^{-1}$) as a function of the rest-frame 
days since the SN explosion. \label{fig:f6}}
\end{figure}

In Figure~\ref{fig:f4} we give an example of a strong (S/N $> 200$) detection on the 
continuum-subtracted image. 
SN~2005kl is the strongest H$\alpha$-emitting object in our sample. 
Its H$\alpha$ luminosity, $L(\mathrm{H}\alpha) \sim 7 \times 10^{41}$ erg~s$^{-1}$
(Table~\ref{table:detect} and Figure~\ref{fig:f5}), 
is about an order of magnitude greater than that of the largest H~II clouds \citep{kenni84},
which strongly suggests that the detected H$\alpha$ emission is (at least partly) 
due to SN-CSM interaction.   

Figure~\ref{fig:f5} shows the plot of the derived H$\alpha$ luminosities as a function of
distance. It is seen that, on average, point sources tend to be brighter than diffuse sources 
(the median luminosities for the two groups are $8 \times 10^{38}$ erg~s$^{-1}$ and 
$2 \times 10^{38}$ erg~s$^{-1}$, respectively), 
but both types of sources can be found at all luminosity levels, except for
the exceptionally bright point source SN~2005kl. 

Further hints of the possible presence of ejecta-CSM interaction may be gained from the
temporal variability of the detected H$\alpha$ sources. SN-CSM interaction usually
produces variable H$\alpha$ emission on the 
timescale of $\sim 100$ days \citep{mau12, fra14}, although there
are known cases, e.g. SN~1988Z, a strongly interacting Type IIn \citep{aret99}) showing 
nearly constant H$\alpha$ luminosity over $\sim 500$ days. Nevertheless, we imaged 38 SNe on at
least two epochs separated by at least 90 days to detect any variability in the net H$\alpha$
emission from the SN site. 

In Figure~\ref{fig:f6} we plot the light curves of 10 detected H$\alpha$ emitters that were observed 
at least three times. It is seen that 6 of them do not show any change that exceeds the uncertainties, 
although SN~1981B, 2013dk and 2014C exhibited continuous variation (either increase or decrease)
during our observations. Flux changes exceeding the photometric errors significantly (i.e. larger
than $3 \sigma$) were found in the case of SN~1985F (Ib, $3.7 \sigma$), 2005kl (Ic, $8.1 \sigma$), 2010gi 
(IIb, $3.3 \sigma$) and 2012fh (Ic, $5.6 \sigma$). SN~2010gi, a Type IIb SN, is a diffuse source (Figure~\ref{fig:f3}) that is more affected by the changing atmospheric conditions, because the
applied flux-scaling method that corrects for changing atmospheric transparency (see above) 
works reliably only for point sources. Thus, we do not believe that the variability detected at 
the position of SN~2010gi is real, and, instead, we attribute this to an instrumental effect. 
The other three objects are SNe-Ibc, and they are all point sources, thus, their variability
is more convincing.
Again, variability of the H$\alpha$ line flux alone cannot be considered as proof of
the ejecta-CSM interaction, but it might give additional support for this hypothesis,
for example in the case of the extremely strong H$\alpha$ emitter SN 2005kl, which is also 
a strongly variable source. 

\section{Discussion}\label{sec:discuss}

Even though it is difficult to prove the existence of ejecta-CSM interaction from 
only narrow-band imaging, without spectroscopic and/or multi-wavelength (X-ray and/or
radio) observations, our data at hand strongly suggest that CSM interaction is
the likely explanation for the H$\alpha$ emission in at least a few of the 
SNe in our sample. The collision between the SN blast wave and the expelled,
slowly moving H-envelope can certainly produce H$\alpha$ emission in the
same order of magnitude as observed, as already demonstrated by SN~2001em
(Section~1). 

SN~2004dk (Type Ib) is another example, where SN-CSM interaction was detected in 
X-rays by {\it XMM-Newton} shortly after explosion \citep{pooley07}, 
and by radio observations at late
phases \citep{stock09, wellons12}. Although \citet{wellons12} concluded that "it is unlikely
that the shockwave is interacting with a H-rich common envelope" based on a single
nebular spectum of SN~2004dk that showed only a weak, narrow, unresolved H$\alpha$ emission 
feature attributed to ISM \citep{maeda08}, our solid (S/N $> 50$) detection of a strong point source at the 
SN position may suggest that the blast wave has finally reached the expelled H-rich envelope. 
The lack of variability (Fig.\ref{fig:f6}), however, does not strengthen this interpretation, 
thus, follow-up observations will provide important constraints for this case.

\begin{figure}[ht!]
\figurenum{7}
\includegraphics[width=8.5cm]{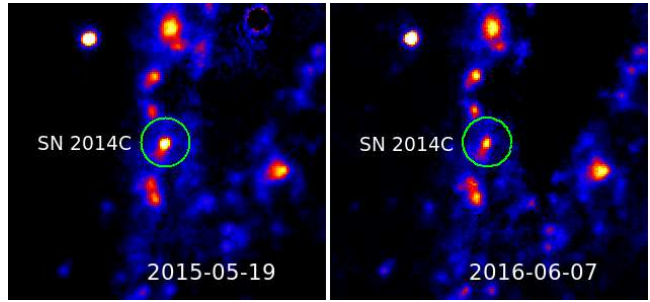}
\caption{ Detection of continuum-subtracted H$\alpha$ emission from SN 2014C (encircled) 
in May 2015 (at $\sim 500$ days after explosion) and June, 2016 (at $\sim 900$ days).
The decrease of the H$\alpha$ flux between the two epochs is apparent. The size of the frames
is $1 \times 1$ arcmin$^2$.
\label{fig:f7}}
\end{figure}

A more recent example occurred during the progress of our project.
After emerging from solar conjunction, the Type Ib SN~2014C started to
show strong H$\alpha$ emission $\sim 110$ days after explosion, even
though H was absent from its spectrum during the photospheric phase
\citep{danny16}. Accompanied by contemporaneous strong X-ray and radio 
emission, this remarkable metamorphosis was explained by ejecta-CSM
interaction \citep{raf16}, the same mechanism for which we search.

We detected the appearance of the H$\alpha$ emission from SN~2014C 
on our DIAFI frames taken between 2015 May and 2016 June 
($\sim 500$ and $\sim 900$ days after explosion, respectively).  
Figure~\ref{fig:f7} shows the continuum-subtracted DIAFI frames obtained
on 2015-05-19 and 2016-06-07 (left and right panels, respectively), with the
object encircled. Its H$\alpha$ luminosity (Table~\ref{table:detect}) is
in agreement with the independent spectroscopic measurements given by 
\citet{danny16}. This source seems to show continuously decreasing 
net (i.e. continuum-subtracted) H$\alpha$ flux (see Figure~\ref{fig:f6}),
but due to the relatively large errorbars of its photometry, the detected 
overall flux change did not exceed $2 \sigma$ during the $\sim 1$ year baseline
of our observations.

More recently, SN~2014C was also detected in mid-IR bands with 
the {\it Spitzer} Space Telescope by the SPIRITS project, as a variable 
point source \citep{tinya16}.  
Note that \citet{tinya16} observed an {\it increase} of the mid-IR fluxes
from SN~2014C between 200 and 600 days after explosion, while our observations
span only the later ($t > 500$ days) phases. 
The latest {\it Spitzer} observations from the SPIRITS project, partly published by \citet{tinya16},
show a slow but continuous decline of mid-IR fluxes between $\sim 600$ and $\sim 1000$ days
in both the $3.6\mu$ and the $4.5\mu$ bands, in agreement with our H$\alpha$ observations.
 
\citet{tinya16} explained 
the observed re-brightening of SN~2014C in the mid-IR as due to shock heating (either
radiative or collisional) of pre-existing dust around SN~2014C, which is
in good agreement with the observations in other bands, including our
H$\alpha$ detections. This highlights the importance of simultaneous
observations in non-optical bands, in order to detect additional tracers
of the ejecta-CSM interaction that may discriminate between the different
mechanisms producing the contemporaneous H$\alpha$ emission. 

The relatively low number of detected H$\alpha$-emitting point sources (13 out of 99 SNe, 
i.e. $\sim 13$ percent) is more-or-less consistent with the recent estimate for the rate of the
SN~2001em/2014C-like events published by \citet{raf16} based on late-time re-brightening in the
radio. From 41 Type Ibc SNe having radio coverage at $t > 500$ days after explosion, 
\citet{raf16} found 4 cases (SN~2001em, SN~2003gk, SN~2007bg and SN~iPTF11qcj, $\sim 10$ percent of the sample) 
when luminous radio re-brightening was observed years after core collapse. Since radio light 
curves are usually attributed to SN-CSM interaction, the emergence of radio fluxes at such late phases are 
very strong signs for the ejecta-CSM collision.  Their result ($\sim 10$ \%) is consistent with our 
detection rate ($\sim 13$ \%) when taking into account that at least some of our detected H$\alpha$ 
emitters could also be compact H~II areas very close to the SN site, i.e. the detected emission may
not be due to ejecta-CSM interaction in every case.

Variability in the detected H$\alpha$ line emission can be another tracer for the SN-CSM interaction, 
because H~II clouds are usually close to ionization/recombination equilibrium, and do not show 
noticeable variation in their emission line strengths. On the contrary, existing models on the 
ejecta-CSM interaction \citep[e.g.][]{cc06, vanmarle10} 
predict several forms of variability in the produced radiation. 
The most commonly accepted scenario is the close link between the X-rays produced by
the RS and the H$\alpha$ line emission coming from the H-rich dense shell in between
the FS and the RS (see Section~1). Since the main source of the escaping H$\alpha$ line
photons are the X-rays absorbed by the shell, any kind of variation in the absorbed X-rays 
can imply change in the emergent H$\alpha$ flux. At the beginning of the interaction the
models predict a relatively quick rise ($\sim 500$ days) of the X-ray flux \citep{cc06}
followed by a slower decline after the FS passed through the dense shell and the shell
got accelerated by the SN ejecta piling up from behind \citep{cl89}. The deceleration of
the RS, ${\rm v_{RS}} \sim t^{-1/(n-2)}$, where $n$ is the power-law index of the density profile
in the outer part of the SN ejecta, is also partly responsible for the decrease of the
H$\alpha$ luminosity since $L_X \sim {\rm v_{RS}}^3$ \citep{cf03, nymark06} and 
$L_{\mathrm{H}\alpha} \sim \eta L_X M_S^{9/40}$, where $\eta \sim 0.1$ is the efficiency of converting
the X-ray flux to H$\alpha$ photons and $M_S$ is the mass of the dense CSM shell \citep{cc06}. 
In addition, at later phases the shell
gets diluted by expansion, reducing the X-ray optical depth, which causes further decrease in the
produced H$\alpha$ radiation. 

In Figure~\ref{fig:f6} there are examples for increasing (e.g. SN~1981B, 2012fh) as well as 
decreasing (SN~2005kl, SN~2014C) H$\alpha$ luminosities. In the simple interaction scenario described above, the
increasing H$\alpha$ fluxes might be connected with an early-phase, still strengthening ejecta-CSM
interaction, while the declining H$\alpha$ fluxes could be explained by a more evolved interaction.
The possibility for the developing ejecta-CSM interaction around the Type Ia SN~1981B, as suggested by 
the persistent (even though not statistically significant) increase in its H$\alpha$ luminosity, 
is especially interesting, since is the only SN~Ia in our sample that shows such a phenomenon.
Since the baseline of our survey is still less than 1000 days, and we have only a few measured points
for each SNe, it is premature to draw any definite conclusion on such details. Future observations
will be useful to decide whether these SNe are indeed subject to ejecta-CSM interaction.

\section{Conclusions}\label{sec:concl}

Our narrow-band imaging survey of old, H-deficient (Type Ibc, Ia and IIb) SNe 
through H$\alpha$ filters resulted in the following:
\begin{itemize}
\item{detection of continuum-subtracted H$\alpha$ emission from 27 SNe sites (see Table~\ref{table:detect}), 13 of which are point sources at our resolution;}
\item{detection of significant variation (exceeding $3 \sigma$) of the H$\alpha$ emission
from 3 SNe: SN~1985F, SN~2005kl, SN~2012fh (all SN~Ibc and point sources). SN~2010gi, a diffuse source
around a SN~IIb, also shows some sort of variability, but its reality is more uncertain; }
\item{a strong, variable H$\alpha$ emitter, SN~2005kl (SN~Ic), for which the 
H$\alpha$ luminosity exceeds that of the typical H~II regions by an order of magnitude;}
\item{detection of H$\alpha$ emission from the known late-time interacting SN~2004dk and SN~2014C; }
\item{possible variation in the H$\alpha$ emission from SN~1981B (Ia), SN~2012P (IIb), 
SN~2013dk (Ic) and SN~2014C (Ib).}
\end{itemize}

We conclude that in the case of the three variable H$\alpha$ emitters, i.e. 
SN~1985F, SN~2005kl and SN~2012fh, the source of the net H$\alpha$ emission is likely 
the ongoing ejecta-CSM interaction. 
The robustness of our methodology is demonstrated by the successful detection 
of the well-known interacting SN~2004dk and SN~2014C. 
The number of the detected point-source H$\alpha$ emitters, 13 out of 99 SNe, is consistent with the
recently estimated rate ($\sim 10$ percent) of such events \citep{raf16}.

In addition to continuing our narrow-band {\bf imaging} for all nearby Type~I SNe, we plan to obtain spectroscopic and multi-wavelength data, from radio to X-ray bands, for the SNe listed in 
Table~\ref{table:detect} to confirm and characterize the ejecta-CSM interaction.  
We are also investigating the archival radio and X-ray coverage of these SNe 
(Pooley et al., in prep).

\acknowledgments
This work was supported in part by the following grants: 
NSF Fellowship AST-1302771 (JMS), NSF Grant AST-1109881 (JCW), 
NKFIH/OTKA grants PD-112325 (TS) and NN-107637 (JV) 
of the Hungarian National Research, Development and Innovation Office. 
We acknowledge the helpful
assistance of the McDonald Observatory staff during our observing runs.
This research has made use of NASA's Astrophysics Data System Service 
operated by the Smithsonian Astrophysical Observatory (SAO), and the NASA/IPAC 
Extragalactic Database (NED) 
which is operated by the Jet Propulsion Laboratory, California Institute 
of Technology, under contract with the National Aeronautics and Space Administration. 

\vspace{5mm}
{\vskip 6pt{\large \it Facilities}: McDonald Observatory: 2.7m (DIAFI)}

{\vskip 6pt{\large \it Software}: IRAF, WCStools, HOTPANTS, SExtractor, YODA}


\end{document}